\documentstyle[aps,floats,epsf]{revtex}  

\draft
\begin{document}

\twocolumn[\hsize\textwidth\columnwidth\hsize\csname
@twocolumnfalse\endcsname

\title{The virtual crystal approximation revisited:\\
Application to dielectric and piezoelectric properties of perovskites}
 
\author{L. Bellaiche$^{1}$ and David Vanderbilt$^{2}$}

\address{$^{1}$ Physics Department,\\
                University of Arkansas, Fayetteville, Arkansas 72701, USA\\
          $^{2}$  Center for Materials Theory,
                   Department of Physics and Astronomy,\\
         Rutgers University, Piscataway, New Jersey 08855-0849, USA}

\date{August 24, 1999}

\maketitle

\begin{abstract}
We present an approach to the implementation of the virtual
crystal approximation (VCA) for the study of properties of solid
solutions in the context of density-functional methods.  Our
approach can easily be applied to any type of pseudopotential,
and also has the advantage that it can be used to obtain
estimates of the atomic forces that would arise if the real atoms
were present, thus giving insight into the expected displacements
in the real alloy.  We have applied this VCA technique within the
Vanderbilt ultrasoft-pseudopotential scheme to predict dielectric
and piezoelectric properties of the Pb(Zr$_{0.5}$Ti$_{0.5}$)O$_{3}$
solid solution in its paraelectric and ferroelectric phases,
respectively.  Comparison with calculations performed on ordered
alloy supercells and with data on parents compounds demonstrates
the adequacy of using the VCA for perovskite solid solutions.  In
particular, the VCA approach reproduces the anomalous Born
effective charges and the large value of the piezoelectric
coefficients.
\end{abstract}

\pacs{PACS numbers: 71.23.-k, 71.15.Hx, 77.84.Dy, 71.15.Mb}

\vskip2pc]

\narrowtext

\section{Introduction}

The application of first-principles electronic band-structure
methods to the study of disordered alloys and solid solutions
requires some approximation for the treatment of the alloy
disorder.  A ``direct'' approach is to make use of the supercell
approximation, i.e., to study one or more disordered configurations
in a supercell with artificially imposed periodic
boundary conditions.  Such calculations generally require
the use of very large supercells in order to mimic the distribution
of local chemical environments, and tend to be computationally
very demanding.  A much simpler and computationally less expensive
approach is to employ the virtual crystal approximation (VCA)
\cite{VCA}, in which one studies a crystal with the primitive
periodicity, but composed of fictitious ``virtual'' atoms that
interpolate between the behavior of the atoms in the parent
compounds.  This technique has seen wide use in band-structure
calculations
\cite{Gironcoli,Marzari,Marco,Papa,Pickett,Slavenburg,%
APL97,Chen,LWW,Ramer}.
Another possible approach would be to make use of the coherent
potential approximation (CPA) \cite{CPA}, but unfortunately
the CPA is generally
not well suited for use in first-principles total-energy methods.
A different way to go beyond the VCA is to carry out
a systematic perturbation expansion in the difference between
the true and VCA potentials, an approach that is sometimes
referred to as ``computational alchemy'' \cite{Gironcoli,Marzari,Marco}.
However, this method is much more complicated than the
usual VCA, requiring the use of density-functional linear-response
techniques.

Clearly the VCA has the advantages of simplicity and computational
efficiency, if two possible concerns can be addressed.  First and
foremost is the question of the accuracy of the VCA approximation.
Previous work has demonstrated good accuracy for the VCA in some
semiconductor and ferromagnetic materials
\cite{Gironcoli,Marzari,Marco,Papa,Pickett,Slavenburg},
but it was found to be inadequate for an accurate treatment of the
electronic structure of some unusual semiconductor systems
\cite{APL97,Chen,LWW}.  Until the recent pioneering work of Ramer
and Rappe \cite{Ramer}, nothing was known about the ability of the VCA
to describe the properties of an important class of materials,
the ferroelectric perovskite solid solutions.
Their work strongly suggests that these alloys are good candidates
for modeling with the VCA, since it reproduces the strain-induced
transitions of ordered supercells of Pb(Zr$_{0.5}$Ti$_{0.5})$O$_{3}$.
However, it is not known whether the VCA is good enough to predict
the anomalous dielectric and piezoelectric properties of perovskite
solid solutions.

A second concern is more technical.  By its nature, the VCA is
closely tied to the pseudopotential approximation.  Indeed, unless
pseudopotentials are used, it is hopeless to apply the VCA to the
usual case of isoelectronic substitution (i.e., atoms belonging to
the same column but different rows of the Periodic Table).
However, as pseudopotential methods have matured, it has become
less obvious what is the correct or optimal way to implement the VCA.
For the case of local pseudopotentials, the implementation is
straightforward \cite{APL97}: the potential of the virtual system
made from the (A$_{1-x}$B$_{x}$)C alloy is generated simply
by compositionally averaging the potentials of the parent AC and
BC compounds,
\begin{equation}
V_{\rm VCA} ({\bf r}) = (1-x) V_{\rm AC} ({\bf r}) + x V_{\rm BC} ({\bf r}) ~. 
\end{equation}
In practice this is usually done in Fourier space by averaging
$V_{\rm AC} ({\bf G})$ and $V_{\rm BC} ({\bf G})$.
In the case of semilocal (e.g, Hamann-Schl\"uter-Chiang\cite{HSC})
pseudopotentials, a similar averaging of the radial potentials
$V_{{\rm A},l}(r)$ and $V_{{\rm B},l}(r)$ can be done separately
in each angular momentum channel $l$. 
 However, with the fully
non-local Kleinman-Bylander type separable pseudopotentials
\cite{KB} that are most commonly used in the current generation of
electronic-structure calculations, the implementation of the VCA is
neither straightforward nor unique.  For example, Ramer and Rappe
\cite{Ramer} discuss four different ways of implementing the VCA 
for such
pseudopotentials, each of them providing different physical
results.  Similarly, the best way of applying the VCA to the case
of ultrasoft pseudopotentials \cite{USPP} is less obvious still.

The purpose of the present work is to report progress in addressing
both of the above concerns.  Taking them in reverse order, we first
present a first-principles VCA approach which is easily implemented
for any type of pseudopotential.  The method is demonstrated
and tested in the context of calculations on Pb(Zr$_x$Ti$_{1-x})$O$_3$
(PZT), an important perovskite solid solution.  Our approach also
has the advantage that it can easily be used to obtain estimates of
the atomic forces that would arise if the real atoms were present,
thus giving insight into the expected displacements in the real
alloy.  Such information, which for example is highly relevant to
many properties of ferroelectric systems, is not provided by the
usual VCA techniques.

Second, we use our new approach to evaluate the quality of the
VCA approximation for predicting the unusual dielectric and
piezoelectric properties in ferroelectric perovskite alloys.
Perovskite compounds are known to exhibit anomalous dielectric
properties.  For example, they display anomalously large values of
the Born effective charges, resulting from hybridization between
the transition-metal $d$ and oxygen $2p$ orbitals
\cite{Zhong,Posternak}.  Similarly, piezoelectric coefficients are
large, compared to other classes of materials, both because of the
large Born effective charges and because of the large microscopic
reaction of the internal atomic coordinates to macroscopic strain
\cite{Szabo1,SzaboPRL,LaurentDavid1}.  Using Pb(Zr$_x$Ti$_{1-x})$O$_3$
with $x$=0.5 for our test system, Born effective charges and
piezoelectric coefficients are calculated using our new VCA
approach together with the modern theory of polarization
\cite{Domenic2,Resta}.  We find that the VCA
can be used with fair confidence to predict dielectric and
piezoelectric properties of Pb(Zr$_{0.5}$Ti$_{0.5})$O$_3$ alloys.
As a matter of fact, our VCA technique yields large Born effective
charges that are very nearly equal to the average between the
effective charges of the parent compounds.
Furthermore, comparison with calculations performed on ordered 
Pb(Zr$_{0.5}$Ti$_{0.5})$O$_3$  supercells demonstrates the ability of
the VCA to mimic piezoelectric coefficients
of perovskite solid solutions.

The paper is organized as follows.  In Sec.~II, we implement
our VCA approach in the context of density-functional theory,
emphasizing the advantages of the new approach.  Section III
reports the predictions of this new VCA technique for the Born
effective charges and piezoelectric coefficients of the
Pb(Zr$_{0.5}$Ti$_{0.5})$O$_3$ solid solution in its paraelectric
and ferroelectric phases, respectively.  We conclude in Sec.~IV
with a discussion of perspectives and future directions.
The Appendix contains details about the implementation of our VCA
technique within the Vanderbilt ultrasoft-pseudopotential scheme
\cite{USPP}.

\section{The VCA implementation}

Within a pseudopotential approach to density-func\-tional
theory, the total energy of $N_{v}$ valence electrons can be written
in terms of the one-particle wavefunctions $\phi_{i}$ as
\begin{eqnarray}
E_{\rm tot} [ \{ \phi_{i}\}&,&\{ {\bf{R_I}} \}] =
 U( \{{\bf R_I} \}) +
\sum_i \langle \phi_i | -\nabla^{2} + V_{\rm ext} |  \phi_i \rangle
\nonumber \\
&& +\frac{1}{2} \int \int d{\bf r} \, d{\bf r'}
     \frac{n({\bf r})n({\bf r'})}{\bf|r-r'|} + E_{\rm XC}[n]
~,
\end{eqnarray}
where
\begin{equation}
V_{\rm ext}({\bf r,r'}) = \sum_I
   V_{\rm ps}^I({\bf r-R}_I,{\bf r'-R}_I) ~,
\end{equation}
${\bf R}_I$ is the location of the site $I$, and
$V_{\rm ps}^I$ are the pseudopotentials.
Here, $n(\bf{r})$ is the electron density, $E_{\rm XC}$ is the exchange
and correlation energy, and $U( \{ \bf{R_I} \})$ is the ion-ion
interaction energy. 
A {\it local} pseudopotential takes the form
$V_{\rm ext}({\bf r,r'}) = V_{\rm ext}({\bf r})\,\delta({\bf r-r'})$,
while a {\it non-local} pseudopotential is written as a sum of
projectors.

In either case, it is possible to derive a ``VCA'' operator
equation by simply averaging the pseudopotentials of the alloyed elements
on site $I$,
\begin{equation}
   V_{\rm ps}^I({\bf r,r'}) =
    (1-x) V_{\rm ps}^A ({\bf r,r'})
      + x V_{\rm ps}^B ({\bf r,r'}) ~,
\end{equation}
where, e.g., $A$=Ti and $B$=Zr in Pb(Zr$_{x}$Ti$_{1-x}$)O$_{3}$.
For the sites occupied by the non-alloyed $C$ elements
[Pb or O in Pb(Zr$_x$Ti$_{1-x})$O$_3$], one simply takes
\begin{equation}
   V_{\rm ps}^I({\bf r,r'}) =
    V_{\rm ps}^C ({\bf r,r'}) ~.
 \end{equation}
Then $V_{\rm ext}$ can be written
\begin{equation}
V_{\rm ext}({\bf r,r'}) = \sum_I \sum_\alpha
  w^I_{\alpha} V_{\rm ps}^{\alpha}({\bf r-R}_{I\alpha},{\bf r'-R}_{I\alpha}) ~,
\label{eq:vext}
\end{equation}
where $V_{\rm ps}^{\alpha}$ is the pseudopotential for an atom of
type $\alpha$ and $w^I_\alpha$ is a ``weight'' which specifies the
statistical composition on site $I$.  In cubic Pb(Zr$_x$Ti$_{1-x})$O$_3$,
for example, we set
$w=1$ for Pb and $w=0$ otherwise at the cube-corner site;
$w=1$ for oxygen and $w=0$ otherwise at the three face-center sites;
and $w=x$ and $w=1-x$ respectively for Zr and Ti (and zero
otherwise) at the cube-center site.
%
%
In other words, we think of this crystal as composed of {\it six}
atoms in the primitive cell: the usual one Pb and three oxygen
atoms, and two ``ghost'' atoms (Zr and Ti) sharing the {\it same}
lattice site (and having weights $x$ and $1-x$ respectively).
We
then treat this unit cell containing six ``atoms'' in the usual
first-principles pseudopotential approach, solving the Kohn-Sham
equations in the presence of the potential given by
Eq.~(\ref{eq:vext}).

This approach has the advantage of requiring only very slight
modifications to the usual first-principles pseudopotential code.
One simply inputs the weight $w$, along with the position and
atom type, of each ``atom'' in the unit cell.  These weights
are then used in just a few places, e.g., in the construction of
the total external potential (\ref{eq:vext}).  Perhaps the only
subtlety is in the treatment of the Ewald energy \cite{Ewald} in the ion-ion
interaction term $U$ appearing in Eq.~(2).  Here, we clearly have
to prevent any Coulomb interaction between two ``ghost atoms''
on the same site, or else the Ewald energy would be infinite.
In fact, the Ewald energy that we calculate is that of a crystal
having valence charge
\begin{equation}
  \overline{z}_v^I = \sum_\alpha w^I_{\alpha} ~ z_v^{\alpha}
\label{eq:zv}
\end{equation}
on site $I$ (so that, in the case of isoelectronic substitution,
this is just the usual Ewald energy).  In practice, as long as the
same gaussian width is used for all species when splitting the
real-space and reciprocal-space Ewald contributions, this can
be done very simply by replacing $z_{v}^{\alpha}$ $\rightarrow$
$w_{\alpha} z_{v}^{\alpha}$ inside the program and deleting the
infinite on-site interaction terms that would occur in the
real-space Ewald sum.

There are three definite advantages to this new VCA approach.
First of all, it is extremely easy to implement, as already indicated;
only the minor modifications of Eqs.~(\ref{eq:vext}) and
(\ref{eq:zv}) have to be implemented when starting from a conventional
first-principles pseudopotential code.  Second, in contrast to the
approach of Ref.~\cite{Ramer}, there is no need to generate a
pseudopotential for each virtual atom.  Here, the pseudotentials
are created once and for all for each true atomic species; only the
weights w$_I$ change when dealing with a new composition of the
solid solution.  Third, the alloyed elements are still considered as
separate atomic species (with corresponding weights), rather than
creating a single virtual atom as a whole.

This last point is less trivial than it might seem.  Because the
two ``ghost atoms'' on a site are considered as separate ``atoms,''
one can consider responses to the displacement of {\it just one}
of these ``atoms'' alone.  In fact, the program automatically
reports the forces on all the ``atoms'' in the unit cell, including
those on the two ghost atoms separately.  For structures of low
symmetry, these forces need not be the same \cite{explan-a}.
These forces can provide some hints about the atomic distortions
that would occur in the true disordered materials.  The {\it
magnitudes} of the force differences also act as a kind of internal
diagnostic for the appropriateness of using the VCA for the system
of interest.  A large magnitude would suggest that the VCA
approximation is not expected to be very accurate, while a smaller
value implies that the VCA can be used with fair confidence to
mimic structural properties of the disordered alloy under
consideration.

It is also straightforward, using our approach, to use
finite-difference methods to evaluate other kinds of response to
the separate ghost-atom displacements.  For example, the Born
effective charges can be defined as the first-order polarization
changes with respect to first-order sublattice displacements.  As
we will see in Sec.~III, the contribution of each ghost atom to
the Born effective charge of the whole virtual atom can thus easily
be calculated.  To do the finite-difference calculation, we simply
compute the change in polarization as the two ghost atoms are
displaced to slightly different positions.  Of course, to be
meaningful, any real physical quantity (e.g., the derivative of
some observable with respect to displacement) ultimately has to be
evaluated at the configuration of identical ghost-atom positions.

The discussion above has been limited to norm-conserving
pseudopotentials, but the extension to the case of Vanderbilt
ultrasoft pseudopotentials \cite{USPP} is fairly straightforward.
This extension is discussed in the Appendix.  In fact, all of our
tests presented below have been carried out within the ultrasoft
formulation.

\section{Application to perovskite solid solutions}

\subsection{Born effective charges of the paraelectric
Pb(Zr$_{0.5}$Ti$_{0.5}$)O$_{3}$ alloy}

Our first goal is to determine the dynamical effective
charges of the Pb(Zr$_{0.50}$Ti$_{0.50}$)O$_{3}$ solid solution in its
paraelectric phase, as predicted by the VCA. This alloy is usually denoted
as PZT. For this purpose, we first perform local-density
approximation (LDA) \cite{LDA} calculations
within the Vanderbilt ultrasoft-pseudopotential scheme
on the cubic perovskite structure, using our VCA technique. As
detailed in Ref.~\cite{Domenic}, a conjugate-gradient technique is
used to minimize the Kohn-Sham energy functional.  The Pb $5d$, Pb
$6s$, Pb $6p$, Zr $4s$, Zr $4p$, Zr $4d$, Zr $5s$, Ti $3s$, Ti
$3p$, Ti $3d$, Ti $4s$, O $2s$ and O $2p$ electrons are treated as
valence electrons. A weight $w$ of $1$ is assigned to Pb and
oxygen atoms on their corresponding sites, while $w=0.5$ for both
Ti and Zr at the cube-center site [see Eqs.~(6) and (7)].
Consequently, the VCA calculation includes 44
electrons per cell.  We use the Ceperley-Alder exchange and
correlation \cite{Ceperley} as parameterized by Perdew and Zunger
\cite{Perdew}.  A (6,6,6) Monkhorst-Pack mesh \cite{Monkhorst} is
used in order to provide converged results \cite{Domenic}. The
lattice parameter a$_{0}$ is fully optimized by minimizing the
total energy, and is found to be very well described by Vegard's law.
In other words, a$_{0}$ is very nearly equal to the compositional
average between the lattice constants of pure PbTiO$_3$ and pure
PbZrO$_3$ given in Ref.~\cite{Domenic}.

To mimic the paraelectric phase, the atoms are kept in the ideal
cubic positions.
The dynamical effective charge Z$_{33,\alpha}^{*}$ of each real and ghost 
atom is then calculated by
using the formula 
\begin{equation}
 dP_z = \sum_\alpha w_\alpha \, Z^*_{33,\alpha} \, du_{z,\alpha}~~~~~~~~,
\end{equation}
where $dP_z$ is the change in polarization along the $z$-direction induced
by the  displacements $du_{z,\alpha}$ of the $\alpha$ atoms along the
z-direction, and $w_\alpha$ refers to the weight assigned to the atoms.   
We allow the two ghost atoms to be at different
atomic positions in order to compute the contribution
of each of them on the Born effective charge of the whole virtual atom.
We follow the procedure introduced in Ref.\cite{Domenic2} which
consists in directly calculating the spontaneous polarization as a
Berry phase of the Bloch states.  Technically, we use roughly 660
Bloch states to assure a good convergence of the effective
charges.

\begin{table}
\caption{VCA dynamical effective charges $Z_{33}^{*}$ of
Pb(Zr$_{0.5}$Ti$_{0.5}$O$_{3}$), as compared with those of the
parent compounds and their average.
Last two rows show the comparison of the individual VCA
$Z_{33}^{*}$ values for Zr and Ti with those of the parent
compound [16].}
\label{Table I}
\begin{tabular}{ldddd}
Atom &VCA  & PbZrO$_3$ & PbTiO$_3$ & Ave\\
\tableline
Pb      &    3.92 &  3.92 &   3.90 &  3.91 \\
$\langle B\rangle$
        &    6.47 & 5.85 & 7.06 & 6.46\\
O$_{1}$ & $-$2.54 &  $-$2.48 & $-$2.56 & $-$2.52 \\
O$_{3}$ & $-$5.28 &  $-$4.81 &  $-$5.83 & $-$5.32 \\
\tableline
$B$: Zr &    9.62 & 5.86 & ---  & ---  \\
$B$: Ti &    3.32 & ---  & 7.06 & ---  \\
\end{tabular}
\end{table}

Table I shows the Born effective charges of the different atoms, as
well as the compositional average between the $Z_{33}^{*}$ of pure
paraelectric PbTiO$_3$ and PbZrO$_3$ as given in Ref.~\cite{Zhong}.
In this Table, the averaged transition-metal atom interpolating
between Zr and Ti is referred to as $\langle B\rangle$, and the oxygen atoms are
grouped into two kinds: those denoted O$_{3}$, located between two
$\langle B\rangle$ atoms along the $z$ direction; and those denoted O$_{1}$,
located between two $\langle B\rangle$ atoms in the perpendicular directions
\cite{SzaboPRL,Alberto0}.  One can first see that the VCA reproduces
very well the effective charges of Pb, $\langle B\rangle$ and O atoms
found in paraelectric PZT solid solutions
  \cite{LaurentJorgeDavid,LaurentJorgeDavid2}: 
$Z^{*}_{33}$ is around 4, $-$2.5, and $-$5.5 for Pb, O$_{1}$ and
O$_{3}$ atoms, respectively, while the effective charge of the
$\langle B\rangle$ atom is close to 6.5.  In fact, Table I shows that the VCA
approximation essentially averages the effective charges of the parent
compounds for Pb, O$_{1}$, O$_{3}$ and $\langle B\rangle$.  
The VCA approach is thus able to mimic the weak and subtle
interaction between the $d$ orbitals of the transition-metal atom,
treated as a single atom, and the O $2p$ orbitals. This interaction
is responsible for the anomalous effective charges of both O$_{3}$
and  $\langle B\rangle$ atoms \cite{Zhong}.
On the other hand, Table I demonstrates that the $Z^*$ contributions
of the Zr or Ti ghost atoms in PbZr$_{0.5}$Ti$_{0.5}$O$_3$ are
{\it not} well approximated by the $Z^*$'s of the corresponding
Zr or Ti atom in the parent (PbZrO$_3$ and PbTiO$_3$) compounds.
This can be understood by
realizing that the distance between Zr and O atoms in our VCA
simulation of PbZr$_{0.5}$Ti$_{0.5}$O$_{3}$ is smaller than the
corresponding distance in  PbZrO$_{3}$. The hybridization between
the $d$ orbitals of the Zr ghost atom and the O $2p$ orbitals then
differs from the corresponding hybridization in PbZrO$_{3}$. This
leads to  an enhancement of the effective charge of the Zr ghost
atom in PbZr$_{0.5}$Ti$_{0.5}$O$_{3}$ with respect to the
effective charge of Zr in PbZrO$_{3}$.  Inversely, the effective
charge of the Ti ghost atom in PbZr$_{0.5}$Ti$_{0.5}$O$_{3}$ is
much smaller than the effective charge of Ti in
PbTiO$_{3}$, since the distance between Ti and O atoms in the VCA
calculation of PbZr$_{0.5}$Ti$_{0.5}$O$_{3}$ is larger than the
corresponding distance in  PbTiO$_{3}$.  Interestingly, the
underestimation for Ti cancels with the overestimation for Zr,
yielding an effective charge of the whole $\langle B\rangle$ atom which is
nearly equal to the average between the effective charges of the B atoms in
the parents compounds.

\subsection{Piezoelectric coefficient of the ferroelectric
Pb(Zr$_{0.5}$Ti$_{0.5}$)O$_{3}$ alloy}

\begin{table}
\caption{Structural parameters of Pb(Zr$_{0.5}$Ti$_{0.5}$O$_{3}$)
within our VCA approach (denoted `VCA') and for the
supercell ordered along the [100] direction used in Ref.~[20]
(denoted `Ordered 1'). $\Delta z$ are the ferroelectric atomic
displacements, in $c$-lattice units.}
\label{Table II}
\begin{tabular}{ldd}
~~           &~~VCA  &~~Ordered 1 \\
\tableline
~~$\Delta z (Pb)$   &~~$-$0.0486  &~~$-$0.0480 \\
~~$\Delta z (\langle B\rangle)$    &~~+0.0076  &~~+0.0064 \\
~~$\Delta z (O_1)$  &~~+0.0790  &~~+0.0827 \\
~~$\Delta z (O_3)$  &~~+0.0585  &~~+0.0555 \\
\end{tabular}
\end{table}

We now apply the VCA procedure to determine the piezoelectric
coefficients e$_{ij}$ of the ferroelectric tetragonal  P4mm ground
state of the Pb(Zr$_{0.50}$Ti$_{0.50}$)O$_{3}$ solid solution as predicted by
the VCA.  We use here the lattice constants minimizing the total energy
of the ordered supercell of Ref.~\cite{LaurentDavid1}; that is, the
lattice parameter a$_{0}$ and the tetragonal axial ratio c/a are
equal to 3.99 \AA~ and 1.0345, respectively.
The atomic displacements are fully optimized by minimizing the total energy and
the Hellmann-Feynman forces, the latter being smaller than 0.05
eV/\AA~ at convergence. During these minimizations, the ghost atoms 
can move but always share the same position. The Hellmann-Feynman force
on the virtual $\langle B\rangle$ atom is simply the sum of
the forces on the Zr and Ti
``ghost'' atoms, i.e., having  a weight of 0.5 in Eqs.~(6) and (7).  In
the ferroelectric VCA ground state of PZT, the force on the Ti atom
is along the polarization direction, i.e., along the z-axis. This
force is exactly opposite to the force on the Zr atom.
The magnitude of these forces is found to be 1.1 eV/\AA, which is
less than three times larger than the force used to get convergent
results in Ref.~\cite{Singh}.  This indicates that the VCA approach
can be used with some confidence to describe the properties of PZT
alloys.  As a matter of fact, Table II shows that the VCA can
reproduce remarkably well
the atomic displacements leading to  the appearance of
ferroelectricity in the ordered Pb(Zr$_{0.50}$Ti$_{0.50}$)O$_{3}$ alloy
\cite{LaurentDavid1}.

Once the ferroelectric ground state is determined, the modern
theory of polarization \cite{Domenic2,Resta} is used to calculate
the piezoelectric coefficient of PZT within our VCA procedure.
More precisely, the piezoelectric coefficients e$_{ij}$ can be
computed via \cite{David2}
\begin{equation}
e_{ij}= \frac{1}{2\pi\Omega} \sum_{\alpha} R_{\alpha, i}
\frac{d}{d\eta_j} (\Omega {\bf G_{\alpha}} \cdot {\bf P} ) ~~~~~~~,
\end{equation}
where $\Omega$ is the cell volume and $\alpha = {1,2,3}$ runs
over the three real-space lattice vectors ${\bf R_{\alpha}}$ and
reciprocal lattice vectors  ${\bf G_{\alpha}}$, and
$\eta_{j}$ is the macroscopic strain. 
Eq.~(9) has recently been
derived in order to make the piezoelectric coefficients independent
of the choice of branch of the Berry phase \cite{David2}. At the
same time, Eq.~(9) automatically eliminates of the so-called
``improper'' terms \cite{SzaboPRL} as required to correctly
predict the piezoelectric coefficients \cite{David2}.  Technically,
Eq.~(9) is evaluated by finite differences between two strained
configurations: first that of the ferroelectric ground state, and
then for an additional 1\% strain relative to this ground state.
In the second run, the relative atomic coordinates naturally
have to be reoptimized in response to the applied strain.

As done in Ref.~\cite{Szabo1,SzaboPRL,LaurentDavid1,Dalcorso}, the
piezoelectric coefficients can be decomposed into ``clamped-ion''
and ``internal-strain'' contributions,
\begin{equation}
e_{33}=e_{33,c}~+~e_{33,i}~~~~~~~.
\end{equation}
The ``clamped-ion'' or ``homogeneous-strain'' contribution
$e_{33,c}$ is given by Eq.~(9) for the case $i=j=3$ and {\it
evaluated at vanishing internal strain} (that is, {\it without}
allowing the additional relaxation of the relative atomic
coordinates that would be induced by the strain).  $e_{33,c}$
reflects electronic effects, measuring the extent to which
the Wannier centers fail to follow the homogeneous strain.  The
``internal-strain'' part $e_{33,i}$ measures just those
contributions to the piezoelectric response coming from
internal distortions, i.e., reflecting the extent to which
the ions fail to follow the homogeneous strain.  In practice,
$e_{33}$ and $e_{33,c}$ are computed, and $e_{33,i}$ is then
obtained from their difference.

\begin{table}
\caption{Piezoelectric coefficients in C/m$^{2}$ of \protect\linebreak
Pb(Zr$_{0.5}$Ti$_{0.5}$O$_{3}$) within our VCA approach (denoted
`VCA'), for the supercell ordered along the [100]
direction used in Ref.~[20] (denoted `Ordered 1'), for the
supercell ordered along the [001] direction used in Ref.~[18]
(denoted `Ordered 2'), and for the supercell ordered along
the [111] direction used in Ref.~[18] (denoted `Ordered 3'). }
\label{Table III}
\begin{tabular}{ldddd}
~~           &~~VCA  &~~Ordered 1 &~~Ordered 2  &~~Ordered 3  \\
\tableline
~~e$_{33}$   &~~4.4  &~~3.4      &~~4.8       &~~3.6\\
~~e$_{33,c}$ &~~$-$0.8 &~~-0.8     &~~$-$0.7      &~~$-$0.7 \\
~~e$_{33,i}$ &~~5.2  &~4.2       &~~5.4       &~~4.3\\
\end{tabular}
\end{table}

The results for e$_{33}$, e$_{33,c}$ and e$_{33,i}$, as predicted by
the VCA, are shown in Table III and compared with various
calculations on ordered supercells.  One can first notice that the VCA
is able to reproduce not only the magnitudes but also the signs of the
piezoelectric coefficients of ordered supercells.  In particular,
the clamped-ion contribution, which is negative and independent of
the ordering, is very well described by the VCA approximation.
Similarly, the VCA results for e$_{33,i}$  and e$_{33}$ lie between
those of the different ordered supercells, for which a larger
spread exists.  Overall, the results shown in Tables II and III
confirm the adequacy of the VCA to mimic ferroelectric properties of
PZT.

\section{Conclusions}

In summary, we have developed a new first-principles virtual crystal
approach. This method  (1) is easy to implement, (2) does not require
the generation of pseudopotentials for each alloy composition,
and (3) its outputs,
via the computation of the Hellmann-Feynman forces on the ``ghost''
alloyed elements, provide a hint about the ability of the VCA to mimic
properties of the disordered alloys under consideration.  This
technique has been applied, within the Vanderbilt
ultrasoft-pseudopotential scheme
\cite{USPP}, to predict dielectric and piezoelectric
properties of the Pb(Zr$_{0.5}$Ti$_{0.5}$)O$_{3}$ solid solution in its
paraelectric and ferroelectric phase, respectively.  Comparison
with calculations performed on ordered supercells and with data on parent
compounds demonstrates the adequacy of using the VCA for such properties,
at least for isoelectronic perosvkite solid solutions. More work is
needed to assess the ability of VCA to describe properties of
heterovalent alloys \cite{LaurentJorgeDavid,LaurentDavid2}, i.e.,
systems in which the alloyed elements belong to different columns of
the periodic table (e.g., Pb(Sc$_{0.5}$Nb$_{0.5}$)O$_{3}$
or Pb(Mg$_{1/3}$Nb$_{2/3}$)O$_{3}$).

The present study also strongly suggests that
first-principles derived effective-Hamiltonian methods, already
available for simple perovskite systems
\cite{ZhongDavid,Rabe,Krakaeur,Jorge,Alberto1,Alberto2,Cockayne},
can be used with confidence to predict finite-temperature
properties of perovskite (isoelectronic) solid solutions,
by modeling these alloys within the VCA approach.

\section{Acknowledgments}  

L.B.~thanks the financial assistance provided by the
Arkansas Science and Technology
Authority (grant N99-B-21), and P.~Thibado for the loan of a computer. 
D.V.\ acknowledges the financial support of
Office of Naval Research grant N00014-97-1-0048. 
N.J.~Ramer and A.M.~Rappe are also thanked for communicating their results
\cite{Ramer} to us prior to publication. Cray C90 computer time
was provided at the NAVO MSRC under ONRDC \# 1437.

\section{Appendix: implementation of the VCA approach within
the Vanderbilt ultrasoft-pseudopotential scheme}

In this appendix, we indicate how Eq.~(6) can be realized when using the
Vanderbilt's ultrasoft pseudopotentials scheme \cite{USPP}.
In this approach, the total energy
of $N_{v}$ valence electrons described by the monoelectronic
wavefunctions $\phi_{i}$ is given by

\begin{eqnarray}
E_{\rm tot} [ \{ \phi_{i}\}&,&\{ {\bf{R_I}} \}] =
\sum_i \langle \phi_i | -\nabla^{2} + V_{\rm ext} |\phi_i\rangle \nonumber \\
&& +~~\frac{1}{2} \int \int d{\bf r} \, d{\bf r'}
     \frac{n({\bf r})n({\bf r'})}{\bf|r-r'|} ~~+~~ E_{\rm XC}[n]
 \nonumber \\
&& +~~ U( \{{\bf R_I} \})~~~~~~~,
\end{eqnarray}
where 
the external potential $V_{\rm ext}$ contains a local part
$V_{\rm loc}^{\rm ion}$ and a fully nonlocal part V$_{\rm NL}$,
\begin{eqnarray}
V_{\rm ext} = V_{\rm loc}^{\rm ion} + V_{\rm NL}~~~~~~~.
\end{eqnarray}
The local part contains local ionic contributions
\begin{eqnarray}
V_{\rm loc}^{\rm ion} ({\bf r}) = \sum_I V_{\rm loc}^{\rm ion, I} (|{\bf r-R}_I|)~~~~~~~,
\end{eqnarray}
while the fully nonlocal part is given by 
\begin{eqnarray}
V_{\rm NL} = \sum_{\it{nm},I} D_{\it{nm},I}^{(0)} | \beta_{\it{n}}^{I} \rangle
    \langle \beta_{\it{m}}^{I} |~~~~~~~.
\end{eqnarray}
The functions $\beta_{\it{n}}^{I}$ as well as the coefficients
$ D_{\it{nm},I}^{(0)}$ characterize the pseudopotentials, and thus differ
for different atomic species.

The electron density in Eq.~(10) is given by
\begin{eqnarray}
n(\bf{r}) = \sum_{\it i} \left[ | \phi_{\it i} ({\bf r}) | ^2+\sum_{{\it nm},I}
Q_{\it{nm}}^{I}({\bf r})
 \langle  \phi_{\it i} | \beta_{\it n}^I \rangle \langle
  \beta_{\it m}^I | \phi_{\it i} \rangle\right]
\end{eqnarray}
where the augmentation functions $Q_{\it{nm}}^{I}({\bf r})$ are also
provided by the pseudopotentials and are strictly localized in the
core regions.  The ultrasoft pseudopotential is thus fully
determined by the functions $V_{\rm loc}^{{\rm ion},I}$, $Q_{\it{nm}}^{I}$ and
$\beta_{\it{n}}^{I}$, and by the scalar $D_{\it{nm},I}^{(0)}$.  The algorithm
used to generate these quantities is described in Refs
\cite{USPP,Laasonen}.  The wavefunctions $\phi_{\it i}$ are
eigensolutions of:
\begin{eqnarray}
H | \phi_{\it i}\rangle= \epsilon_{\it i} S | \phi_{\it i} \rangle~~~~~~~,
\end{eqnarray}
where $S$ is an hermitian overlap operator given by
\begin{eqnarray}
S = 1 + \sum_{\it{nm},I} q_{\it{nm}}^I | \beta_{\it n}^I \rangle
 \langle   \beta_{\it m}^I |
\end{eqnarray}
with $ q_{\it{nm}}^I = \int d {\bf r} Q_{\it nm}^{I}({\bf r})$,
and where 
\begin{eqnarray}
H =  -\nabla^{2} + V_{\rm eff} + \sum_{\it{nm},I} D_{\it{nm}}^I
  | \beta_{\it{n}}^{I}\rangle \langle \beta_{\it{m}}^{I} |~~~~~~~.
\end{eqnarray}
Here $V_{\rm eff}$ is the screened effective local potential
\begin{eqnarray}
V_{\rm eff} ({\bf r}) = \frac {\delta E_{\rm tot}} {\delta n({\bf r})}
=  V_{\rm loc}^{\rm ion} ({\bf r})
+ \int d{\bf r'} \frac{n({\bf r'})}{|{\bf r}-{\bf r'}|}
+ \mu_{\rm xc} ({\bf r})
\end{eqnarray}
with $\mu_{\rm xc}=\delta E_{\rm XC}[n] / \delta n(\bf{r})$, and
\begin{eqnarray}
D_{\it{nm}}^I  = D_{\it{nm}}^{(0)} + \int d{\bf r} V_{\rm eff}({\bf r})
   Q_{\it{nm}}^{I}({\bf r}) ~~~~~~~.
\end{eqnarray}

Eq.~(6), which is the fundamental equation underlying our VCA approach,
can thus be realized by simply replacing three ionic quantities provided
by the ultrasoft pseudopotentials, namely $V_{\rm loc}^{\rm ion, \alpha}$,
$D_{\rm nm,\alpha}^{(0)}$, and $Q_{\rm nm}^{\alpha}$,
by their product with the corresponding atomic weight $w_{\alpha}$:
 
 \begin{eqnarray}
 V_{\rm loc}^{ion, \alpha} \rightarrow w_{\alpha} ~ V_{\rm loc}^{ion, \alpha}~~~~~~~,
 \end{eqnarray}  
 
 \begin{eqnarray}
 D_{\rm nm,\alpha}^{(0)} \rightarrow w_{\alpha} ~ D_{\rm nm,\alpha}^{(0)} ~~~~~~~,
 \end{eqnarray}
 
 \begin{eqnarray}
 Q_{\rm nm}^{\alpha} \rightarrow w_{\alpha} ~ Q_{\rm nm}^{\alpha} ~~~~~~~.
 \end{eqnarray}

\newpage

\end{document}